# Gravitationally coupled Quantum Harmonic Oscillator

Pramod Pandya
Department of Information Systems
California State University, Fullerton, CA 92834, USA

We present a quantum harmonic oscillator model of a collapsed star trapped in the potential well of its gravitational field. The model incorporates quantum matter (quantum fields) as a source to classical gravitational field. We describe the gravitationally collapsed star as "a quantum harmonic oscillator in an analogy with an electron trapped in a potential well. The subtle point we raise in this paper is whether a standard black hole scenario is a correct description for a gravitational collapse in a semi-classical framework.

1. **Introduction**

The framework of classical General Relativity predicts formation of astrophysical black holes if certain physical conditions are satisfied by a star under gravitational collapse. In such a scenario, an ultimate collapse to a singular state is unavoidable. Of course, this scenario can be avoided, if space-time local to the singular state is modeled within the framework of semi-classical gravitational theory [1]. In our model we allow for the background space-time to be dynamic rather than fixed, that is, the space-time is non-stationary. Non-stationary space-time does present extreme conceptual problems as regards to unique interpretations of ground state of the quantum fields. A fair amount research has been carried out in the past where quantum fields propagate on a fixed background space-time. In this paper we propose a gravitationally coupled quantum harmonic oscillator, which represents a phase transitioned state of a gravitationally collapsed star.

Gravitationally collapsed star evolves to a pure state (black hole) from its initial mixed (statistical) state. So we consider this phase (pure state) to be unstable, and thus a possible decay to a mixed state. The proposed model treats the background space-time to be non-flat and non-stationary. Hawking's discovery of evaporating black holes ([2],[3]) and how the laws of black hole mechanics equated with the laws of thermodynamics, necessitates a careful analysis of coupling of classical space-time and second-quantized matter field. Particle-creation from a time-dependent gravitational field are cited in ([4],[5]), and the reaction back on the background space-time defines what is known as the back-reaction problem [6].

## 2. The Bosonic matter fields

The dynamics of the collapsing star is described by second quantized matter fields $\hat{\Phi}(x)$, and that the space-time is represented by a classical metric tensor $g_{\mu\nu}(x)$. Therefore, the quantum (scalar) fields act as the actual source of the classical background. In other words, the reaction back on the gravitational field (space-time geometry) caused by the production of scalar particles is incorporated as part of dynamics. To achieve this, it is necessary to include the energy-momentum tensor of the quantized scalar matter-field as the right-hand side of Einstein's gravitational field equations, and then one has to solve this equation in a self-consistently for the metric. This would define the back-reaction problem, and thus leading to more appropriate understanding of the resulting space-time geometry.

Einstein's field equation of classical space-time,

$$G_{\mu\nu}[g_{\mu\nu}(x)] = \kappa T_{\mu\nu}[\hat{\Phi}(x), g_{\mu\nu}(x)]$$

is not correct as stated, since it equates an operator-value to a classical-number. Hence, we consider a semi-classical gravity model, where the dynamics of the space-time geometry is determined by the quantum fields. The obvious modification is to rewrite as,

$$G_{\mu\nu}[g_{\mu\nu}(x)] = <\hat{T}_{\mu\nu}[\hat{\Phi}(x), g_{\mu\nu}(x)]>_\eta$$

where the right-hand side of the above equation is treated as an expectation value of $\hat{T}_{\mu\nu}$ (the energy-momentum tensor acting as a source of the classical gravitational field) with respect to some suitably defined mixed statistical (probably non-equilibrium) states of the system, rather than the vacuum state of the system. We will address the nature of mixed statistical states $\eta$, later on in the paper.

The geometries determined by the modified equation would depend on the mixed statistical states, and the possible geometries ought to be prescribed by a probability distribution. It is evident that, there are many a possible geometries as solutions of the modified equation. The significant of the preceding logic is that a quantum black hole is just one such geometry, from a possible many geometries. In here we note, that to compute expectation value of $\hat{T}_{\mu\nu}$, one has to solve the field equations, and thus determine the statistical states (mixed), which too depend on $g_{\mu\nu}(x)$. In other words, these statistical states will almost certainly depend on the metric tensor, $g_{\mu\nu}(x)$, (think of it as generally covariant-looking version of Gibbs ensemble). A self-consistent solution of the geometry of the space-time and the quantum fields is known as semi-classical gravity. Clearly, the theory is more non-linear than is evident, and the full implications of this approach are cited in the following references ([7,[8],[9],[10],[11],[12],[13]).

The most interesting results, which have been achieved so far using this approach, are from [14],[15],[16] where the authors consider a massive quantized scalar field, but in which the metric tensor is restricted to be Robertson-Walker form. The result of their calculations is an explicit from for the function R(t) which possesses the remarkable feature that the system does not exhibit the classical gravitational collapse but rather 'bounces off' the singularity at R = 0 with the radius R(t) achieving a minimum of the Compton wavelength of the massive scalar particles. Nariai and Tomia [17], have studied modified Lagrangian for the gravitational field, containing terms quadratic in the Ricci tensor and the scalar curvature. For certain choices of parameters in their modified Einstein equations, Nariai and Tomita obtained isotropic, homogeneous cosmological solutions in which the cosmological singularity is replaced by finite minimum in the radius function. Their solutions asymptotically approach Friedmann universe in one time direction, but apparently not in both.

In our paper, we assume a general class of space-time where quantum fields propagate, and seek a solution, which satisfies simultaneously the Einstein's field equations for the space-time, and the field equations for the quantum fields. We consider a real, neutral, scalar quantum field (self-interaction terms are absent) propagating in a time-dependent space-time. We choose the metric tensor $g_{\mu\nu}(x)$ to be diagonal and specified by the function $f^2(t)$, which is not explicitly stated. The differential equations (1), and (2) are derived in section 3 of this paper.

(1) $\quad \dfrac{d^2 z}{d\zeta^2} + a^6(\zeta)\omega^2(\zeta)z(\zeta) = 0 \qquad\qquad -\infty < \zeta < \infty$

(2) $\quad \dfrac{d^2 z}{d\zeta^2} + a^6 \omega^2 z(\zeta) = 0 \qquad\qquad -\infty < \zeta < \infty$

The reduced equation (2) is a second order linear differential equation with constant coefficients and is obtained from equation (1). As a matter of fact, equations (1), and (2) both resemble that of a harmonic oscillator without the external force. In such a case the solution of equation would be expressed in terms of trigonometric functions. Both equations set up a framework within which we can discuss the nature of quantum black hole. We refer to a similar paper by Wu and Cai [18].

The organization of this paper is as follows. In section 3, the line element is chosen with $g_{\mu\nu}(x)$ to be diagonal, and the resulting Klein-Gordon equation is separated into spatial and time-dependent part. In section 3, the time-dependent part of the Klein-Gordon equation is manipulated and transformed to eigenvalue equation in section 4. In section 5, we present conclusion and suggest future direction the research should follow.

3. **Klein-Gordon Equation**

The Lagrangian density of real scalar quantum field $\phi(x)$ propagating in non-stationary space-time is given as

$$L = \dfrac{\sqrt{-g(x)}}{2}\{g_{\mu\nu}(x)\partial^\nu \phi(x)\partial^\mu(x) + m^2 \phi^2(x)\}$$

The field equation derived from the above Lagrangian density is

$$\partial^\mu \{\sqrt{-g(x)} g_{\mu\nu}(x) \partial^\nu \phi(x)\} = \sqrt{-g(x)} m^2 \phi(x)$$

We choose $g_{\mu\nu}(x)$ to be diagonal,

$$g_{\mu\nu}(x) = \begin{pmatrix} -1 & 0 & 0 & 0 \\ 0 & f^2(t) & 0 & 0 \\ 0 & 0 & f^2(t) & 0 \\ 0 & 0 & 0 & f^2(t) \end{pmatrix}$$

With the choice of coordinates, the field equation is restated as below:

$$\therefore -3\left(\frac{df}{dt}\right)\left(\frac{\partial \phi}{\partial t}\right) - f(t)\left(\frac{\partial^2 \phi}{\partial t^2}\right) + \frac{1}{f(t)}\nabla^2 \phi = m^2 f(t)\phi$$

Since we have a linear differential equation in ϕ(x), we choose

$$\phi(t, \underline{x}) = T(t)X(\underline{x})$$

thus, transform the above differential equation into two equations.

$$\left(\frac{d^2 X(\underline{x})}{dx^2}\right) = -K^2 X(\underline{x})$$

$$f(t)\left(\frac{d^2 T(t)}{dt^2}\right) + 3\left(\frac{df}{dt}\right)\left(\frac{dT}{dt}\right) + [m^2 f(t) + K^2]T(t) = 0$$

4. **Derivation of eigenvalue equation.**

Next we transform the time-dependent equation derived in section 3, such that the coefficient of the first derivative of T(t) is zero. For this work we refer to Tricomi [19]. (Appendix for detailed calculations.)

$$f(t)\left(\frac{d^2 T}{dt^2}\right) + \frac{3}{2}\left(\frac{df}{dt}\right)\left(\frac{dT}{dt}\right) + [f(t)m^2 + K^2]T(t) = 0$$

put,

$$A(t) = f(t), \quad B(t) = \frac{3}{2}\left(\frac{df}{dt}\right), \quad C(t) = [f(t)m^2 + K^2]$$

and,

$$\phi = \frac{1}{\alpha^2(t)} \exp\left[-\frac{3}{2} \int \frac{df/dt}{f(t)} dt\right]$$

$$= \frac{1}{\alpha^2(t)} f^{-\frac{3}{2}}(t)$$

$$T = \alpha(t)z$$

$$\zeta = \int \frac{1}{\alpha^2(t)} f^{-\frac{3}{2}}(t) \, dt$$

$$C_1 = A(t)\left(\frac{d^2\alpha}{dt^2}\right) + B(t)\left(\frac{d\alpha}{dt}\right) + C(t)\alpha(t)$$

$$\therefore \ C_1 = f(t)\left(\frac{d^2\alpha}{dt^2}\right) + \frac{3}{2}\left(\frac{df}{dt}\right)\left(\frac{d\alpha}{dt}\right) + [f(t)m^2 + K^2]\alpha(t)$$

$$A_1(t) = A(t)\alpha(t) = f(t)\alpha(t)$$

hence,

$$\left(\frac{d^2 Z}{d\zeta^2}\right) + f^2(t)\alpha^3(t)\left[f(t)\left(\frac{d^2\alpha}{dt^2}\right) + \frac{3}{2}\left(\frac{df}{dt}\right)\left(\frac{d\alpha}{dt}\right) + [f(t)m^2 + K^2]\alpha(t)\right]Z(\zeta) = 0$$

If we choose $\alpha$ to be a constant, then

$$\left(\frac{d^2\alpha}{dt^2}\right) = \left(\frac{d\alpha}{dt}\right) = 0$$

thus,

$$\left(\frac{d^2 Z}{d\zeta^2}\right) + f^2\left[m^2 + \frac{K^2}{f}\right]Z(\zeta) = 0$$

$$\zeta = \int f^{-\frac{3}{2}}(t) \, dt, \quad Z = T$$

$$\left(\frac{d^2Z}{d\zeta^2}\right) + f^3(\zeta)\omega_\kappa^2(\zeta)Z(\zeta) = 0$$

where,

$$\omega_\kappa^2 = \left[m^2 + \frac{K^2}{f}\right]$$

put,

$$a^2(\zeta) = f(\zeta)$$

(1) $$\left(\frac{d^2Z}{d\zeta^2}\right) + a^6(\zeta)\omega_\kappa^2(\zeta)Z(\zeta) = 0 \qquad -\infty < \zeta < +\infty$$

with,

$$\omega_\kappa^2(\zeta) = \left[m^2 + \frac{K^2}{a^2(\zeta)}\right]$$

We want to treat the above differential equation as an eigenvalue problem. The eigenvalues are to be identified with $\omega_\kappa$.

We note that the solutions of equation (1) represents the eigenstates of a quantum harmonic oscillator, trapped in its own potential-well. We have a quantum a state that can get into an excited state, and then settle back into an equilibrium state by radiating away the energy. This would imply that the quantum state is a thermodynamic system with parameters such as entropy, and temperature. In here we realize that this entropy depends on $g_{\mu\nu}(x)$, and its implication that there is some intimate connection between entropy and the gravitational field. The quantum state is already in a mixed state, therefore we need to construct the density operator to represent this mixed state. Next the question arises as to what is this entropy, and temperature? We need to setup relativistic field theoretic statistical mechanics to provide the framework to discuss the resulting thermodynamics.

## 5. Conclusion

The eigenvalue equation we derived in section 4, shows that the frequency, $\omega$ is a function of f(t), a component of $g_{\mu\nu}(x)$. The implication of this result is that the density matrix representing the mixed states η (suitably constructed covariant form of Gibb's Ensemble)[23], depends on the choice of the metric tensor, and the metric tensor depends on the choice of the mixed states. Therefore we need a self-consistent solution of the field equations [21], to show that a gravitational collapse does indeed lead to a classical black hole if back reaction correctly factored in the problem formulation.

The collapsing star is represented as thermodynamic system, and hence it has entropy. Once the collapsing state undergoes a phase transition to a quantum state, this quantum state continues to represent the entropy; that is that the newly formed quantum state is still in a mixed state, rather than in a pure state. If the quantum state is in a pure state, then its entropy would be zero, and the collapsing star would evolved to zero entropy state from non-zero entropy state, which would be a non-unitary evolution of a quantum state – unless quantum entropy allows non-unitary evolution [22]. This quantum state is a new phase of stellar collapse, where the quantum matter and gravitational fields have merged together.

Finally we point that, entropy is well defined for a classical and quantum mechanical (non-relativistic) thermodynamic system, but what about when we apply this concept of entropy to a relativistic system. How do we interpret the concept of entropy for a quantum state?
We are aware of the fact that the entropy is not a Lorentz scalar, so how does one calculate entropy for a relativistic system. Since we are considering a quantum state (which is modeled as a quantum oscillator), what does its vacuum state correspond to? Obviously, the next question we pose is, is the vacuum state unique?

Appendix

$$A(x)\left(\frac{d^2y}{dx^2}\right) + B(x)\left(\frac{dy}{dx}\right) + C(x)y(x) = 0$$

Let y(x) = α(x) z, where α(x) is a given function and z is a new unknown variable.

$$\left(\frac{dy}{dx}\right) = \left(\frac{d\alpha}{dx}\right)z + \alpha\left(\frac{dz}{dx}\right)$$

$$\left(\frac{d^2y}{dx^2}\right) = \left(\frac{d^2\alpha}{dx^2}\right)z + \alpha\left(\frac{d^2z}{dx^2}\right) + 2\left(\frac{d\alpha}{dx}\right)\left(\frac{dz}{dx}\right)$$

Then the above equation is transformed to

$$A(x)\alpha(x)\left(\frac{d^2z}{dx^2}\right) + \left[2A(x)\left(\frac{d\alpha}{dx}\right) + B(x)\alpha(x)\right]\left(\frac{dz}{dx}\right)$$
$$+ \left[A(x)\left(\frac{d^2\alpha}{dx^2}\right) + B(x)\left(\frac{d\alpha}{dx}\right) + C(x)\alpha(x)\right]z(x) = 0$$

Next, put

$$A_1(x) = A(x)\alpha(x)$$

$$B_1(x) = \left[2A(x)\left(\frac{d\alpha}{dx}\right) + B(x)\alpha(x)\right]$$

$$C_1(x) = \left[A(x)\left(\frac{d^2\alpha}{dx^2}\right) + B(x)\left(\frac{d\alpha}{dx}\right) + C(x)\alpha(x)\right]$$

hence,

$$A_1(x)\left(\frac{d^2z}{dx^2}\right) + B_1(x)\left(\frac{dz}{dx}\right) + C_1(x)z(x) = 0$$

Change the independent variable x to $\zeta$ by

$$\zeta = \int \phi(x)dx$$

where $\phi(x)$ is an unknown function of x to be determined.

$$z = z[\zeta(x)]$$

$$\left(\frac{dz}{dx}\right) = \left(\frac{dz}{d\zeta}\right)\left(\frac{d\zeta}{dx}\right) = \phi(x)\left(\frac{dz}{d\zeta}\right)$$

$$\left(\frac{d^2z}{dx^2}\right) = \phi^2(x)\left(\frac{d^2z}{d\zeta^2}\right) + \left(\frac{d\phi}{dx}\right)\left(\frac{dz}{d\zeta}\right)$$

hence, the differential equation is transformed to

$$A_1(x)\phi^2(x)\left(\frac{d^2z}{d\zeta^2}\right) + \left[A_1(x)\left(\frac{d\phi}{dx}\right) + B_1(x)\phi(x)\right]\left(\frac{dz}{d\zeta}\right) + C_1(x)z(x) = 0$$

coefficient of (dz/d$\zeta$) vanishes provided

$$A_1(x)\left(\frac{d\phi}{dx}\right) = -B_1(x)\phi(x)$$

i.e. $$\frac{\phi'(x)}{\phi(x)} = -2\left(\frac{d\alpha}{dx}\right)\frac{1}{\alpha(x)} - \frac{B(x)}{A(x)}$$

On integrating,

$$\phi(x) = \frac{1}{\alpha^2(x)} \exp\left[\int \frac{B(x)}{A(x)} dx\right]$$

and the differential equation is reduced to

$$A_1(x)\phi^2(x)\left(\frac{d^2z}{d\zeta^2}\right) + C_1(x)z(x) = 0$$

ϕ(x) can be evaluated once B(x) and A(x) are specified and α(x) has been chosen.